# Quasirandom Rumor Spreading: An Experimental Analysis

Benjamin Doerr    Tobias Friedrich    Marvin Künnemann
Thomas Sauerwald

**Abstract**

We empirically analyze two versions of the well-known "randomized rumor spreading" protocol to disseminate a piece of information in networks. In the classical model, in each round each informed node informs a random neighbor. In the recently proposed quasirandom variant, each node has a (cyclic) list of its neighbors. Once informed, it starts at a random position of the list, but from then on informs its neighbors in the order of the list.

While for sparse random graphs a better performance of the quasirandom model could be proven, all other results show that, independent of the structure of the lists, the same asymptotic performance guarantees hold as for the classical model.

In this work, we compare the two models experimentally. This not only shows that the quasirandom model generally is faster, but also that the runtime is more concentrated around the mean. This is surprising given that much fewer random bits are used in the quasirandom process.

These advantages are also observed in a lossy communication model, where each transmission does not reach its target with a certain probability, and in an asynchronous model, where nodes send at random times drawn from an exponential distribution. We also show that typically the particular structure of the lists has little influence on the efficiency.

## 1 Introduction

We conduct an experimental analysis of *randomized rumor spreading* protocols (also known as *random broadcast* [9] or the *push model* [3, 13]). Starting with one node of a graph having a piece of information ("rumor") unknown to the other nodes, rumor spreading protocols aim at efficiently making this information known to all nodes. They typically proceed in rounds. In the classical model, in each round, each node that already knows the rumor transmits it to a neighbor chosen uniformly at random. If the rumor is unknown to the neighbor, it becomes informed.

Though not a very elaborate protocol, it works very efficiently on many graph classes. For complete graphs, hypercubes and sufficiently dense random graphs, only $\Theta(\log n)$ rounds are required to inform all nodes (where $n$ is the number of nodes of the graph) [9, 12]. We refer to Section 2 for a precise statement of the model and the results.

Recently, three of the authors introduced a quasirandom variant of the randomized rumor spreading model [5]. Here, each node is equipped with a cyclic list of its neighbors. Once informed, each node chooses a single random neighbor. Its first transmission is directed to that neighbor, the next to its successor in the list, and so on. Hence all transmissions of this node are determined by the first, random one. It was then shown that, independent of the choice of the lists, this protocol also needs only $\Theta(\log n)$ rounds for complete graphs, hypercubes and sufficiently dense random graphs.

While it could be shown that the quasirandom model has a superior runtime on a few graph classes like complete trees and sparse random graphs, unfortunately all other existing results only succeed in showing that the runtime of the quasirandom model for complete graphs, hypercubes and not too sparse random graphs is of the same order as the runtime of the fully random model. The difficulty is that the current theoretical methods are mostly too coarse to make constant factors precise, let alone lower order terms. Since these are relevant in a practical application, we use experimental analysis to obtain non-asymptotic results for concrete graphs. They yield the following results.

We shall see, as expected, that the quasirandom model is generally faster. For sparser graphs, which also represent the practically more relevant setting, savings typically are more than ten percent, which is more than what we expected.

Despite the fact that in the quasirandom model a single random choice has a larger influence than in the fully random model, the quasirandom model is more robust in several respects. However, we see that the deviation of the actual broadcast time from the expected value is much smaller in the quasirandom model. For the sparse random graph $G(n,p)$ with $n = 2^{12}$ and $p = \ln(n)/n$, conditioned on being connected, we observe that in 10% of all runs the fully random model needs more than 13 rounds more than the mean value. However, in 99% of all runs the quasirandom model terminates in less than 6 rounds more than the mean (cf. Figure 2(d)).

We also observe robustness against transmission failures. This is again a well-known strength of the fully random model, cf. [7, 9, 13], and one where again the reduced amount of randomness could lead to inferior results for the quasirandom model. However, even in the setting where each transmission has a 50% chance of not reaching the addressee (without notice to the sender) the quasirandom model keeps its lead. For the hypercube with $2^{12}$ vertices, the average broadcast time is 40.41 in the quasirandom and 45.53 in the random model.

We also discuss the question if the order of the list has an influence on the quality of the protocol. The good news given by the theoretical results is that no such choice can lead to a real failure, that is, for the graph classes discussed so far the difference can at most be a constant factor. Our simulations indicate that there are differences, but they are small for most graphs. This is again good news from the view-point of application, since one advantage of the quasirandom model is that one can use an implicitly given list (which should be present at any node to have some means of addressing neighbors).

The remainder of this paper is organized as follows. We first describe the two broadcasting models and known results for them in Section 2. In Section 3 we compare the average runtimes and their concentrations around the means. In Section 4 we provide an explanation for the good performance of the quasirandom model for sparse graphs. The influence of the lists is discussed in Section 5. In Sections 6 and 7 we show that the advantages observed so far for the quasirandom model also hold in the presence of transmission failures and in an asynchronous model.

## 2 Preliminaries

### 2.1 Broadcasting models.

One of the simplest broadcasting protocol is the so-called push model, which we shall also call *fully random model* (or simply *random model*). There, initially only one node of a graph $G = (V, E)$ owns a piece of information (or equivalently, knows a rumor) which is spread iteratively to all other nodes: in each time-step $t = 1, 2, \ldots$ every *informed* node chooses a neighbor uniformly at random, which the piece of information is sent to. We are interested in how many time-steps are required such that all nodes become informed.

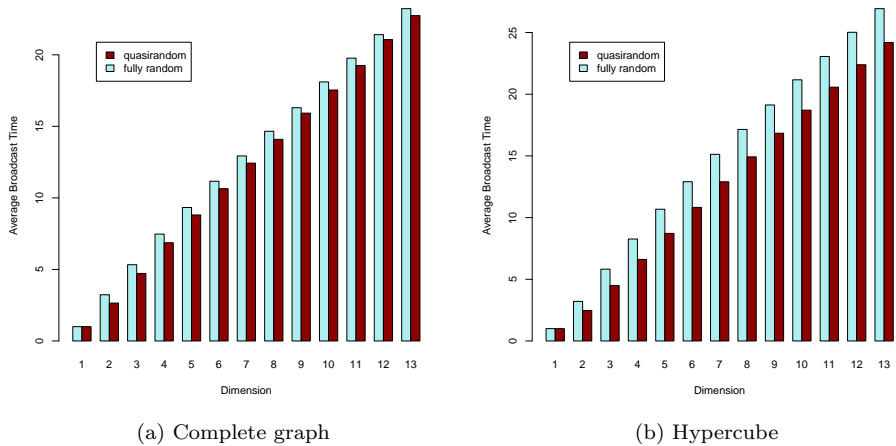

(a) Complete graph  (b) Hypercube

Figure 1: Average broadcast times for complete graphs and hypercubes with $n = 2^1, \ldots, 2^{13}$ nodes.

We compare this random broadcasting model with its quasirandom analogue introduced in [5]. In the quasirandom model, each node has a cyclic list of its neighbors and informs the neighbors in the order of the list. The order of the lists can be arbitrary, but the starting point of each list is assumed to be random. For simplicity, we assume that a node does not stop sending the rumor, even if the list has been completed.

## 2.2 Related work.

The two broadcasting models have been analyzed on different network topologies. For bounded-degree graphs it is known that the random model succeeds in $\Theta(\mathrm{diam}(G))$ with probability $1 - 1/n$ [9], while the quasirandom model succeeds in the same time bound with probability 1 [5].

For the *complete graph* $K_n$ with $n$ vertices where every pair of distinct vertices is adjacent, Frieze and Grimmett [12] showed that $\log_2 n + \ln n + o(\log n)$ rounds suffice with probability $1 - o(1)$. For the quasirandom model it has been shown that there the broadcast time is of order $\Theta(\log n)$ [5] with probability $1 - o(1)$. The time bound was recently improved to $\log_2 n + \ln n + o(\log n)$ by Fountoulakis and Huber [11].

The $d$-dimensional *hypercube* $H_d$ is the graph defined by $V = \{0,1\}^d$ and $E = \{\{u, u(i)\} \mid u \in V, i \in \{1, \ldots, d\}\}$, where $u(i)$ is the bit-vector obtained by flipping the $i$th bit of $u$. For the random model, Feige et al. [9] proved a bound of $\Theta(\log n)$ with probability $1 - 1/n$. Again, the same bound holds for the quasirandom model [5].

The classical *random graph* model introduced by Erdős and Rényi [8] is defined as follows. Let $p \in [0,1]$. Pick an edge between each pair of vertices independently with probability $p$.

If $p \geqslant (1+\varepsilon)\ln(n)/n$, $\varepsilon > 0$, then with probability $1 - o(1)$ the random graph $\mathcal{G}(n,p)$ has the property that the fully random broadcasting protocol is successful in $\Theta(\log n)$ steps with probability $1 - 1/n$ [9]. The same holds for the quasirandom model with the protocol having a success probability of $1 - 1/n$.

For sparse random graphs $\mathcal{G}(n,p)$ with $p = (\ln n + f(n))/n$, where $f(n) \to \infty$ and $f(n) = \mathcal{O}(\log \log n)$, the random model needs at least $\Omega(\log^2 n)$ steps to achieve a success

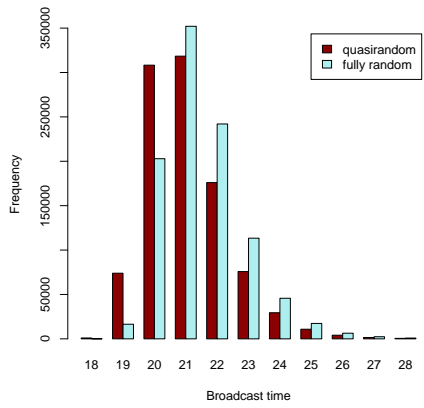
(a) Complete graph $K_{2^{12}}$

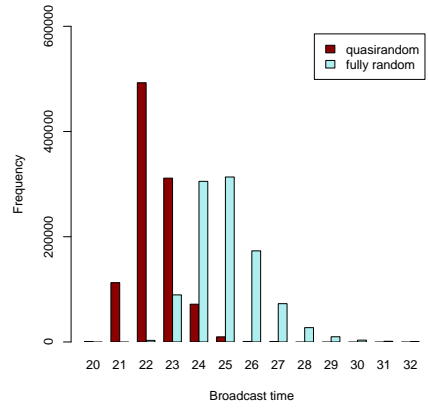
(b) Hypercube $H_{12}$

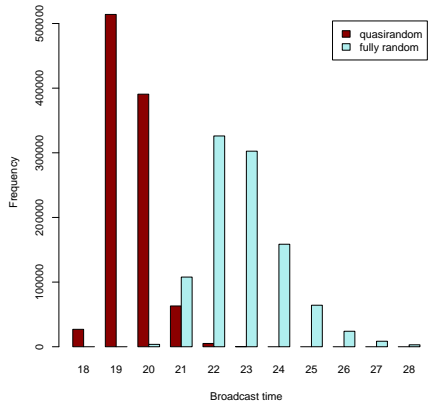
(c) Random 12-regular graphs

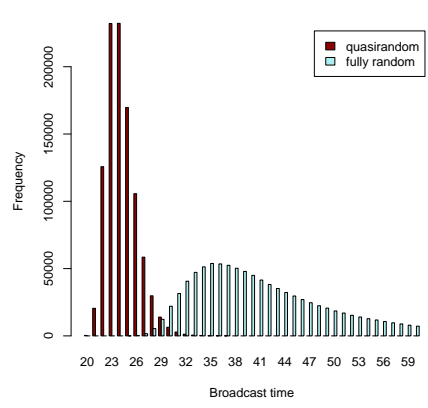
(d) Random graphs with $p = \ln(n)/n$

Figure 2: Empirical distribution of broadcast times on four different graphs with $n = 2^{12}$ nodes.

probability of $1 - 1/n$. Here we observe a provable advantage of the quasirandom model, which needs only $\Theta(\log n)$ rounds to succeed with probability $1 - 1/n$ [5, 6].

Another line of research was conducted by Diks and Pelc [4] and Kim and Chwa [14]. In the language of our quasirandom protocol they construct a list of the neighbors for each vertex irrespective of the starting vertex such that the broadcast time is close to the best deterministically achievable one. They assume that the nodes strictly follow the order of the lists, that is, the first addressee of each vertex is *not* chosen at random, but taken as first entry of the list. Since this deterministic model is quite a different from ours, we do not further refer to it in this work.

### 2.3 Experimental setup.

We use simulations to obtain empirical estimates on the expected broadcast time and on the variance. Unless otherwise stated, we use the following lists also called canonical lists. For complete graphs, every vertex has the same list of neighbors $(1, 2, \ldots, n)$, excluding itself. On hypercubes with dimension $d$, we set the list of each vertex $u$ to $(u(1), u(2), \ldots, u(d))$. For all random graphs the lists contain the neighbors again in increasing order (assuming the vertex set to be $[1..n]$).

The experiments were implemented in C++ with use of the GSL and were run on Sun Fire V20z workstations with AMD Opteron 250 processors. For the seed of the pseudorandom number generator we used the current time. For each experiment, we have performed a large number of repetitions to obtain accurate estimates of the expected broadcast time and the corresponding variance. Each time, the starting vertex $s$ is chosen uniformly at random (unless the graph is vertex-symmetric). For all random graphs, the random and quasirandom model are tested on the same samples.

As generating random graphs in the Erdős-Renyi model can take up to quadratic time, a new sample of this random graph is generated only every thousand runs. Since all experiments were repeated at least 100,000 times and also the starting vertex was chosen uniformly at random in each iteration, we feel that this still gives reliable results.

Since we shall observe that small degree vertices have a large influence on the broadcast times, we also examine *random regular graphs*, see e.g. [18]. We generate the random regular graph uniformly at random with the Steger-Wormald-algorithm [17].

For the case that the random graph generated is not connected, the graph is discarded and a new one is generated.

## 3 Broadcast times

A clear advantage of the quasirandom model is that no node will inform a neighbor a second time (unless it already has informed all its neighbors). This should make the quasirandom model faster, to a stronger extent for sparser graphs than for dense ones. Indeed, as Table 1 shows, we observe considerable gains for sparse graphs and still 2% for the complete graph with $2^{12}$ vertices. Taking into account that informing a neighbor twice does not happen very often in the relatively short runtime, these gains are more than what we expected.

Comparing the results for the hypercube and the random regular graph, we see that even though they both have the same degree distribution, both protocols are faster on the random regular graph and the quasirandom protocol has a stronger advantage here. The latter could be explained by the fact that such random graphs have few cycles only. Hence locally they often look like trees, for which an advantage of the quasirandom protocol could even be proven by theoretical means.

|  | **Random broadcast** | **Quasirandom broadcast** |
|---|---|---|
| Complete graph $K_{2^{12}}$ | $21.50 \pm 1.32$ | $21.04 \pm 1.32$ (=2.1% faster) |
| Hypercube $H_{12}$ | $24.98 \pm 1.32$ | $22.37 \pm 0.82$ (=10.4% faster) |
| Random 12-regular graphs | $22.87 \pm 1.30$ | $19.51 \pm 0.68$ (=14.7% faster) |
| Random graphs $G(n,p)$ with $p = \ln(n)/n$ | $43.20 \pm 11.8$ | $24.28 \pm 1.83$ (=43.8% faster) |

Table 1: Averages and standard deviations of the broadcast times for different graphs with $n = 2^{12}$ vertices.

The highest difference between both models in Table 1 is attained for random graphs $G(n,p)$ with $p = \ln(n)/n$, which is precisely the connectivity threshold for random graphs [8]. We only use the sampled graph if it is connected. We observe empirically a large advantage of 44% for these graphs. In Section 4 we will explain how the large number of small vertices slows down the fully random broadcast on these graphs.

Figure 1 shows the difference between the quasirandom and the random model for different graph sizes, where for each size we have conducted $100,000$ iterations. For complete graphs, one can see a small but stable additive gap between the average observed broadcast times. For hypercubes, this gap seems to be increasing with the dimension.

Finally, we briefly discuss the concentration of the runtime distribution. In addition to the average runtimes, Table 1 gives the empirical standard deviations. For all four examined graph classes, the deviations are higher for the random model. Taking into account that the quasirandom model uses much less independent randomness, this is rather surprising.

As for the averages, the strongest difference occurs for sparse random graphs. There, the standard deviation drops by more than a factor of six. Accordingly, the histogram in Figure 2(d) shows an extremely heavy tail of the fully random broadcast time compared to the quasirandom broadcast time.

## 4 Influence of the degree distribution

In Section 3 we observed that a small minimum degree favored the quasirandom model in comparison to the random one. We now want to further support this explanation.

The largest difference between both models so far is attained for sparse random graphs according to Table 1. From a theoretical point of view, one can show that they have $\omega(1)$ small vertices, i.e., vertices of constant degree. As shown by [2], such small vertices are at distance at least $\Theta(\log n / \log \log n)$ and each neighbor of a small vertex has a degree of $\Theta(\log n)$.

Let us now consider the quasirandom model. By its definition, every small vertex with at least one informed neighbor gets informed within $\mathrm{maxdeg}(G) = \mathcal{O}(\log n)$ further steps with probability 1. On the other hand, consider the random model in a situation where all small vertices have only informed neighbors. Since all these neighbors have degree $\Theta(\log n)$, the expected time to inform a fixed small vertex is $\Theta(\log n)$ and the expected time to inform all $\omega(1)$ small vertices is $\omega(\log n)$. This suggests an asymptotic superiority of the quasirandom model.

On a more concrete level, let us reconsider the sparse random graph with $n = 2^{12}$ at the connectivity threshold $p = \ln(n)/n$. There, the expected number of nodes with degree smaller than five is $\sum_{k=0}^{4} \binom{n-1}{k} p^k (1-p)^{n-k-1} n \approx 339$ and with good probability some of them will have only neighbors with large degree. To inform these, the random protocol has to spend significantly more than the quasirandom protocol. This is illustrated in

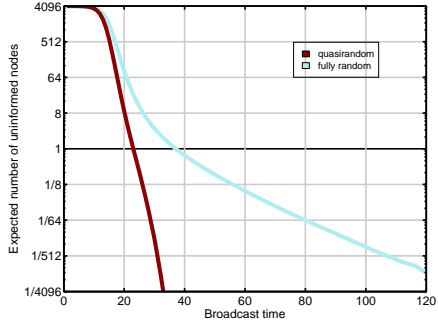

Figure 3: Decrease of the expected number of uninformed nodes during the two broadcasting processes on sparse random graphs with $n = 2^{12}$ nodes and $p = \ln(n)/n$. (Note that the $y$-axis is scaled logarithmically.)

|  | **Random broadcast** | **Quasirandom broadcast** |
|---|---|---|
| Random graphs with $p = \ln(n)/n$ | $43.20 \pm 11.8$ | $24.28 \pm 1.83$ (=43.8% faster) |
| Random graphs with $p = 2\ln(n)/n$ | $26.78 \pm 3.55$ | $22.12 \pm 1.42$ (=17.4% faster) |

Table 2: Averages and standard deviations of the broadcast times for different random graphs with $n = 2^{12}$ vertices.

Figure 3 where one can observe that at the beginning both models do not deviate much. In fact, on average more than 90% of the nodes become informed after 16 or 18 steps in the quasirandom respectively random model. On the other hand, within the 100,000 runs of Figure 3 it took never more than 37 rounds to inform the very last node (typically of very small degree) in the quasirandom model while it took up to 228 rounds for the random model.

This looks much different for denser random graphs. For $p = 2\ln(n)/n$, the expected number of nodes with degree smaller than five is smaller than one. Hence the degree distribution differs significantly from the sparser case and the aforementioned effect diminishes as can be seen in Table 2.

## 5 Influence of the lists

So far in our experiments we only regarded the canonical choice introduced in Section 2 for the cyclic lists describing the orders in which nodes contact their neighbors. Recall that existing theoretical work applies to all kinds of lists. Nevertheless, one might speculate that some orders are more efficient than others, and one might wonder if it is good or not if all nodes use the same lists (excluding of course the node itself as addressee). The results obtained in this section will show that there is some influence, but often it is not very large.

### 5.1 Complete graphs, hypercubes and random graphs

A natural candidate different from the canonical choices and suitable for all graphs are random lists. They are interesting in that they demonstrate what would happen if in the fully random model we were picking the new addressee uniformly at random only from the nodes not contacted so far by the node under consideration. For a practical

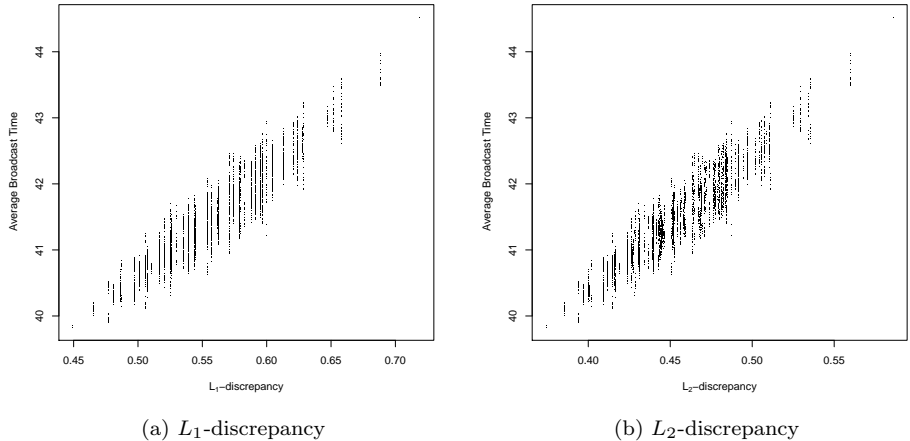

(a) $L_1$-discrepancy  (b) $L_2$-discrepancy

Figure 4: Influence of the lists for the torus graph. For all possible orders of the lists, the plot shows the discrepancy (x-axis) and average broadcast time (y-axis). A good correlation is clearly visible and is confirmed by Pearson product-moment correlation coefficients of $r^2 = 0.869$ and $r^2 = 0.873$ for (a) and (b), respectively.

application, this model suffers slightly from its reduced simplicity.

|  | Random broadcast | Quasirandom broadcast | | |
|---|---|---|---|---|
|  |  | canonic | random | low-discr. |
| Complete graph $K_{2^{12}}$ | $21.50 \pm 1.32$ | $21.04 \pm 1.32$ | $21.48 \pm 1.32$ |  |
| Hypercube $H_{12}$ | $24.98 \pm 1.32$ | $22.37 \pm 0.82$ | $22.32 \pm 0.80$ | $22.36 \pm 0.82$ |
| Random graphs with $p = \ln(n)/n$ | $43.20 \pm 11.8$ | $24.28 \pm 1.83$ | $24.28 \pm 1.82$ |  |
| Torus with $2^{12}$ nodes | $84.09 \pm 2.62$ | $84.57 \pm 2.55$ | $79.15 \pm 2.15$ | $77.10 \pm 1.96$ |

Table 3: The influence of the chosen list on the broadcast times.

An interesting idea for network topologies containing some geometric flavor are low-discrepancy approaches. See e.g. [15, 16] for introductions to some aspects of this broad concept. From the so-far investigated graphs, only the hypercube admits such ideas. Here, instead of informing the neighbors "along the dimension", that is, using the canonical choice $(u(1), u(2), \ldots, u(d))$ as list of vertex $u$, one would serve the dimensions according to a low-discrepancy sequence, that is, take the list $(u(x_1), u(x_2), \ldots, u(x_d))$, where $x_1, \ldots, x_d$ is an injective low-discrepancy sequence in $\{1, \ldots, d\}$ in the sense that each subinterval of the sequence has its entries evenly distributed in the (integer) interval $[1..d]$. There are many constructions of such sequences, cf. again [15, 16]. Taking a van-der-Corput sequence in base two of length 16, rescaling and shifting it to an integer sequence in $[1..16]$ and deleting entries larger than 12 we obtain the low-discrepancy sequence $x = (1, 9, 5, 3, 11, 7, 2, 10, 6, 4, 12, 8)$.

Note that such ideas make little sense for random graphs and complete graphs. First, they both have no regular structure to exploit. In addition, the latter are that symmetric that all sequences (to be used by all nodes) are equivalent.

The results given in Table 3 show that the influence of the order of the lists for complete graphs, hypercube and random graphs is very limited. For the hypercube the use of low-discrepancy sequences is microscopically better that the canonical order, but minimally worse than the use of random sequences. For the complete graph, canonic lists are slightly better than random lists (every node generates its own list). We have no explanation for either of these observations. Having used one million repetitions of each experiment, they seem to be statistically significant, though.

The good news that can be derived from these experiments is that for all these graphs, the particular choice of the lists does change little. In particular, we may use lists implicitly given from the network organization without suffering a loss.

## 5.2 Torus graphs

To trigger some influence of the lists, we now regard the 2-dimensional *torus graph*. Due to its superlogarithmic diameter, it is less relevant as network topology in situations where fast dissemination of information is sought after. However, with its regular and geometric structure, we did expect and do observe a considerable influence of the choice of the lists used.

Let $n$ be a square number. The torus graph has the vertex set $\{0, 1, \ldots, \sqrt{n} - 1\}$. Each vertex $i$ has eight neighbors $\{((i_1 + j_1), (i_2 + j_2)) \mid \|j\|_\infty = 1\}$, where here and in the following all additions are modulo $\sqrt{n}$. The directions these eight neighbors lie in have a canonical (counterclockwise) cyclic order, say $d_1 = (1,0)$, $d_2 = (1,1)$, $d_3 = (0,1)$, and so on. Consequently, the canonic list associated with vertex $u$ is $(u + d_1, u + d_2, \ldots)$. Again, we use a low-discrepancy sequence of length eight, say $x = (1, 5, 3, 7, 2, 6, 4, 8)$, and associate the list $(u + d_{x_1}, u + d_{x_2}, \ldots)$ to each vertex $u$.

Table 3 provides some graphical evidence for a different behavior of these list structures. The canonic lists result in even slightly higher broadcast times than fully random broadcasting, which is significantly beaten by quasirandom broadcasting with random lists and (again significantly better) with the low- discrepancy lists.

In order to provide more convincing evidence if there is a correlation between the discrepancy of the lists and the broadcast times, we measured the broadcast time for *all* possible lists (that is, with all vertices having its lists constructed from the same $x$, for all bijective $x\colon [1..8] \to [1..8]$).

In order to make this computationally tractable, this was performed on the torus graph with $n = 2^{10}$ vertices with 1000 runs for every $x$.

For an $x$ like this, we define its discrepancy as follows. Let $I$ and $J$ be intervals in $[1..8]$, possibly with "wrap-around". Ideally, if $x$ was perfectly evenly distributed, $\{x_i \mid i \in I\}$ should have $|I| \cdot |J|/8$ elements in $J$. Hence the discrepancy for these two intervals is

$$\operatorname{disc}(x, I, J) := \big||\{x_i \mid i \in I\} \cap J| - |I| \cdot |J|/8\big|.$$

For $p > 0$, we define the $L_p$-discrepancy of $x$ by

$$\operatorname{disc}_p(x) := \left(\sum_{I,J} \operatorname{disc}(x, I, J)^p\right)^{1/p},$$

where $I$ and $J$ run over all intervals in $\{1, \ldots, 8\}$ with wrap-around.

Interestingly, we observe a strong correlation between the broadcast time and the discrepancy of the list. Figure 4 shows the average broadcast time relative to the $L_1$- and $L_2$-discrepancy of the list. Every point represents the average broadcast time for one $x$. A correlation is clearly visible and is confirmed by Pearson product-moment correlation coefficients of $r^2 = 0.869$ and $r^2 = 0.873$ for the $L_1$- and $L_2$-discrepancy, respectively.

We also investigated the influence of the list structures on the expansion of the information in the early stage of the process. Figure 5 shows the set of informed vertices

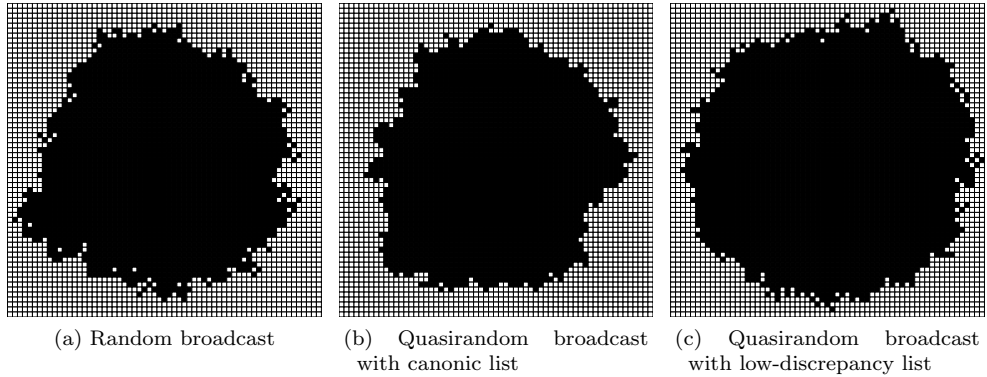

(a) Random broadcast  (b) Quasirandom broadcast with canonic list  (c) Quasirandom broadcast with low-discrepancy list

Figure 5: Informed vertices in a sample run of a torus graph after 50 time steps when the center vertex was informed at the beginning. The number of informed vertices is 2118, 2008, and 2546, respectively. The difference of the radius of the largest inscribed and the smallest circumscribed circle is 10.18, 9.48, and 8.70, respectively.

|  | Random broadcast | Quasirandom broadcast | | |
|---|---|---|---|---|
|  |  | canonic | random | low-discr. |
| Avg. radius difference | $10.25 \pm 1.67$ | $9.57 \pm 1.70$ | $9.03 \pm 1.50$ | $8.96 \pm 1.50$ |
| Avg. number of informed nodes | $2150 \pm 123$ | $2061 \pm 122$ | $2407 \pm 113$ | $2546 \pm 104$ |
| Normalized avg. radius difference | $0.221 \pm 0.005$ | $0.211 \pm 0.005$ | $0.184 \pm 0.004$ | $0.178 \pm 0.004$ |

Table 4: Description of the set of informed vertices if the broadcast process is stopped after 50 time steps on a torus graph with $n = 63^2$ nodes. As expected from Table 3, the average number of informed nodes is highest for the low-discrepancy lists and smallest for canonic lists.

of a torus graph with $n = 63^2$ nodes after 50 time steps when the central vertex was informed at the beginning. It is clearly visible that the number of informed nodes is significantly larger for the quasirandom processes with low-discrepancy lists. The average number of informed nodes after this number of time steps (averaged over 100,000 runs) is shown in Table 4.

The regularity of the dissemination can be measured by comparing the smallest radius such that all informed vertices are at most that far away from the center with the maximum radius such that the corresponding circle around the center contains only informed vertices.

The average radius differences shown in Table 4 indicate a more regular expansion of the set of informed vertices for the quasirandom protocols. This still holds if the radius differences are normalized with fair radius, the radius of the perfect circle containing the average number of informed nodes (last row in Table 4). Such a normalization makes sense, since in particular the low-discrepancy versions of the quasirandom protocol inform a significantly larger area in the same time.

## 6 Robustness against failures

A second important aspect of broadcast protocols, besides their broadcast time, is robustness against transmission errors. A good protocol should still work moderately efficiently even if some transmissions fail or some nodes for whatever reason do not participate as intended.

Naturally, randomized protocols are very robust. In Feige et al. [9], a model was considered where communications links break down forever. For complete graphs it was proven that even if up to $n/3$ links chosen by an adversary break down, randomized broadcast still requires only $\mathcal{O}(\log n)$ steps.

We consider a model where each transmission reaches its destination with some probability $f > 0$ independently. It was shown in [7] that for an arbitrary graph, the runtime of the random model with these transmission failure is at most $6/f$ times the runtime of the random model without failures.

|  | Random broadcast | Quasirandom broadcast |
| --- | --- | --- |
| **Complete graph $K_{2^{12}}$** | | |
| no transmission failures | $21.50 \pm 1.32$ | $21.04 \pm 1.32$ (=2.14% faster) |
| 50% failure chance | $39.38 \pm 3.15$ | $39.29 \pm 3.18$ (=0.23% faster) |
| Increase (multiplicative) | 1.832 | 1.867 |
| **Hypercube $H_{12}$** | | |
| no transmission failures | $24.98 \pm 1.32$ | $22.37 \pm 0.82$ (=10.4% faster) |
| 50% failure chance | $45.53 \pm 3.23$ | $40.41 \pm 2.68$ (=11.2% faster) |
| Increase (multiplicative) | 1.822 | 1.806 |
| **Random 12-regular graphs** | | |
| no transmission failures | $22.87 \pm 1.30$ | $19.51 \pm 0.68$ (=14.7% faster) |
| 50% failure chance | $42.38 \pm 3.20$ | $36.81 \pm 2.67$ (=13.1% faster) |
| Increase (multiplicative) | 1.854 | 1.888 |
| **Random graphs with $p = \ln(n)/n$** | | |
| no transmission failures | $43.20 \pm 11.81$ | $24.28 \pm 1.83$ (=43.8% faster) |
| 50% failure chance | $84.37 \pm 24.34$ | $67.60 \pm 17.89$ (=19.9% faster) |
| Increase (multiplicative) | 1.953 | 2.784 |

Table 5: Average broadcast times for different graph types with a 50% chance of transmissions failing unnoticed by the sender and without transmission failures. Averages are taken over 1,000,000 iterations.

With the fully random model being that robust against different kinds of failures, it is a natural question if the quasirandom broadcast model with its greatly reduced degree of randomness still has comparable robustness properties. To gain some understanding of this issue, we analyze the quite pessimistic model that each transmission fails independently at random with probability one half, and that this remains unnoticed by the sender. Hence the sending node continues his schedule with the next node on his list. Note that in this failure model the advantage that a low-degree node informs all its neighbors in time equal to its degree, clearly vanishes. We would thus expect the superiority of the quasirandom model to reduce significantly.

However, this does not happen. For complete graphs, hypercubes and random regular graphs on $n = 2^{12}$ vertices, we see from Table 5 that both the fully random and the quasirandom model in the presence of 50% transmission failures slow down by factors between 1.81 and 1.89. The standard deviation increases by factors around 3 (and this still is relatively small).

We do not have an explanation for the finding that the quasirandom model compared to the fully random one is sightly more affected by failures in the complete graph topology and slightly less in the hypercube topology. This effect, however, seems real and was observed also for smaller graph sizes with more repetitions of the experiment.

For sparse random graphs, transmission failure seem to have a stronger impact on the quasirandom model, however, it remains significantly faster than the random protocol (by about 20%, still the largest gain in all graph classes regarded).

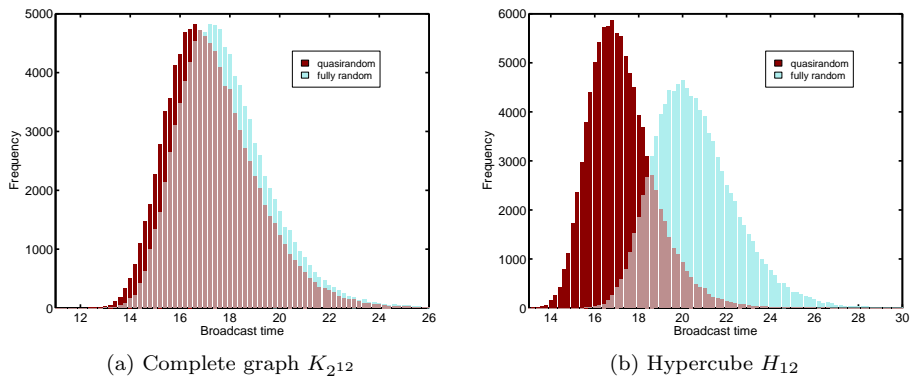

(a) Complete graph $K_{2^{12}}$      (b) Hypercube $H_{12}$

Figure 6: Histograms of the observed runtime over 100,000 runs in the asynchronous setting. The width of each bar is 0.2.

The reason for this stronger impact seems to be that on sparse random graphs, the quasirandom model was exceptionally advantageous. This advantage is partially lost in the lossy model. More specifically, a difficult situation in these sparse random graphs is informing vertices with small degree (e.g., one), that have only neighbors with large degree. Here the quasirandom model wins a lot by the property that a node of degree $d$ surely informs all its neighbors in $d$ rounds. Of course, this property does not exist in the lossy model. Here, the action of a single node resembles more the one it has in the random model.

# 7 Asynchronous broadcasting

The broadcasting model discussed so far made the assumption that the agents (nodes) act in a synchronized manner. This assumption is not completely inline with the idea of a self-organized broadcasting protocol not guided by a central authority. There are works on a continuous model called "Richardson's growth model" in the mathematics community, e.g., [1, 10, 12]. In this model, nodes inform other nodes on a continuous time scale. The order in which a fixed node makes its transmission is the same as in the discrete-time models discussed so far. However, the times to elapse between two consecutive transmissions (and between a node first getting informed and its first transmission) are chosen according to an exponential distribution with expectation one[1]. For this model, a bound of $\Theta(\log n)$ was shown for complete graphs [12] and hypercubes [1, 10].

To see how (a) this model compares to the synchronized one and (b) how the random and quasirandom variants compare in this model, we also implemented it. We performed simulations on the 12-dimensional hypercube and the complete graph, both with $2^{12}$ nodes. Our empirical results averaged over 100,000 iterations are given in Table 6.

Comparing with data on the non-lossy model in Table 5, we see that, generally, the asynchronous model is faster than the synchronized one. The standard deviations are slightly larger. The relative comparison between the different graph classes as well as the two broadcast protocol is similar to the synchronized version. The advantage of the quasirandom protocol is slightly higher for the asynchronous version, though.

---

[1]Actually, in the original "Richardson's growth model", the expectation is $1/\deg(G)$ on regular graphs. However, to make the results of this section comparable to the ones of the synchronous broadcasting, we use an appropriate scaling by setting the expectation to 1.

|  | Random broadcast | Quasirandom broadcast | |
| --- | --- | --- | --- |
| **Complete graph $K_{2^{12}}$** | $17.79 \pm 1.82$ | $17.32 \pm 1.82$ | (=2.6% faster) |
| **Hypercube $H_{12}$** | $20.57 \pm 1.88$ | $17.23 \pm 1.50$ | (=16% faster) |
| **Random 12-regular graphs** | $19.43 \pm 1.86$ | $15.83 \pm 1.46$ | (=18.5% faster) |
| **Random graphs with $p = \ln(n)/n$** | $40.87 \pm 12.32$ | $22.47 \pm 3.10$ | (=45% faster) |

Table 6: Empirical broadcast times for different graph types in the asynchronous model.

## 8  Discussion

In this work, we experimentally analyzed the quasirandom version of the classical randomized rumor spreading model. Our investigation shows a number of interesting facts, which previous work via mathematical means was not able to show. In such, this work nicely complements existing theoretical work. The latter gives a guarantee that the quasirandom model does not fail, and in fact is not worse than the random model, no matter how the neighbor lists are chosen. Note that due to the sheer number of possible lists, there is no way to gain such a worst-case analysis experimentally.

On the other hand, this work indicates that the typical behavior of the quasirandom model is better than what the theoretical bounds show, and also better than the random model. This statement holds in particular with respect to expected runtimes, deviations from the mean value and robustness against transmission failures.

Given the shown differences between the two models, this work also motivates a further development of the methods to analyze dependent randomized algorithms, so that results of this kind can also be obtained in a mathematically rigorous way.

## References

[1] B. Bollobás and Y. Kohayakawa. On Richardson's model on the hypercube. In B. Bollobás and A. Thomason, editors, *Combinatorics, Geometry and Probability*, pages 129–137. Cambridge University Press, 1997.

[2] C. Cooper and A. Frieze. The Cover Time of Random Regular Graphs. *SIAM Journal of Discrete Mathematics*, 18:728–740, 2005.

[3] A. Demers, D. Greene, C. Hauser, W. Irish, J. Larson, S. Shenker, H. Sturgis, D. Swinehart, and D. Terry. Epidemic algorithms for replicated database maintenance. In *6th ACM Symposium on Principles of Distributed Computing (PODC)*, pages 1–12, 1987.

[4] K. Diks and A. Pelc. Broadcasting with universal lists. *Networks*, 27:183–196, 1998.

[5] B. Doerr, T. Friedrich, and T. Sauerwald. Quasirandom Rumor Spreading. In *19th ACM SIAM Symposium on Discrete Algorithms (SODA)*, pages 773–781, 2008.

[6] B. Doerr, T. Friedrich, and T. Sauerwald. Quasirandom rumor spreading: Expanders, push vs. pull, and robustness. In *36th International Colloquium on Automata, Languages and Programming (ICALP)*, volume 5555 of *Lecture Notes in Computer Science*, pages 366–377. Springer, 2009.

[7] R. Elsässer and T. Sauerwald. On the Runtime and Robustness of Randomized

Broadcasting. In 17*th International Symposium on Algorithms and Computation (ISAAC)*, pages 349–358, 2006.

[8] P. Erdős and A. Rényi. On random graphs. *Publ. Math. Debrecen*, 6:290–297, 1959.

[9] U. Feige, D. Peleg, P. Raghavan, and E. Upfal. Randomized broadcast in networks. *Random Structures and Algorithms*, 1:447–460, 1990.

[10] J. Fill and R. Pemantle. Percolation, first-passage percolation and covering times for Richardson's model on the $n$-cube. *Annals of Applied Probability*, 3:593–629, 1993.

[11] N. Fountoulakis and A. Huber. Quasirandom rumor spreading on the complete graph is as fast as randomized rumor spreading. *SIAM J. Discrete Math.*, 23:1964–1991, 2009.

[12] A. Frieze and G. Grimmett. The shortest-path problem for graphs with random arc-lengths. *Discrete Applied Mathematics*, 10:57–77, 1985.

[13] R. Karp, C. Schindelhauer, S. Shenker, and B. Vöcking. Randomized Rumor Spreading. In 41*st IEEE Symposium on Foundations of Computer Science (FOCS)*, pages 565–574, 2000.

[14] J. Kim and K. Chwa. Optimal broadcasting with universal lists based on competitive analysis. *Networks*, 45:224–231, 2005.

[15] J. Matoušek. *Geometric Discrepancy*. Springer-Verlag, Berlin, 1999.

[16] H. Niederreiter. *Random number generation and quasi-Monte Carlo methods*, volume 63 of *CBMS-NSF Regional Conference Series in Applied Mathematics*. Society for Industrial and Applied Mathematics (SIAM), Philadelphia, PA, 1992.

[17] A. Steger and N. C. Wormald. Generating random regular graphs quickly. *Combinatorics, Probability & Computing*, 8:377–396, 1999.

[18] N. C. Wormald. Models of random regular graphs. *Surveys in Combinatorics*, 267:239–298, 1999.